\documentclass[twocolumn,twocolappendix,tighten,resetfootnote]{aastex701}

\usepackage{amsmath}
\usepackage{graphicx}
\usepackage{booktabs}
\usepackage{multirow}
\usepackage{xcolor}

\newcommand{\teff}{T_{\rm eff}}

\newcommand{\mas}{\mathrm{mas}}
\newcommand{\rco}{\theta_{\rm CO}/\theta_{\rm phot}}

\newcommand{\missingfigure}[1]{%
\fbox{\parbox[c][0.23\textheight][c]{0.92\linewidth}{%
\centering Figure file not supplied in the working directory\\[1ex]
\texttt{\detokenize{#1}}}}}
\newcommand{\safeincludegraphics}[2][]{%
\IfFileExists{#2}{\includegraphics[#1]{#2}}{\missingfigure{#2}}}

\newcommand{\UMich}{Department of Astronomy, University of Michigan, Ann Arbor, MI 48109, USA}
\newcommand{\CHARA}{The CHARA Array of Georgia State University, Mount Wilson Observatory, Mount Wilson, CA 91023, USA}

\newcommand{\EXE}{Astrophysics Group, Department of Physics \& Astronomy, University of Exeter, Stocker Road, Exeter, EX4 4QL, UK}

\shorttitle{The Hot, Continuum-compact State of V509 Cas}
\shortauthors{Anugu et al.}

\begin{document}

\title{The Hot, Continuum-compact State of the Yellow Hypergiant
V509 Cassiopeiae: CHARA Diameters, CO-band Angular-extension
Constraints, and Companion Limits}

\correspondingauthor{Narsireddy Anugu}
\email{nanugu@gsu.edu}

\author[0000-0000-0000-0000]{Narsireddy Anugu}
\affiliation{\CHARA}
\email{nanugu@gsu.edu}

\author[0000-0001-5415-9189]{Gail H. Schaefer}
\affiliation{\CHARA}
\email{gschaefer@gsu.edu}

\author[0000-0001-8537-3583]{Douglas R. Gies}
\affiliation{\CHARA}
\email{dgies@gsu.edu}

\author[0000-0002-1823-3975]{Anni Kasikov}
\affiliation{Tartu Observatory, University of Tartu, Observatooriumi 1, T\~oravere 61602, Estonia}
\email{anni.kasikov@ut.ee}

\author[0000-0002-3380-3307]{John D. Monnier}
\affiliation{\UMich}
\email{monnier@umich.edu}

\author[0000-0001-6017-8773]{Stefan Kraus}
\affiliation{\EXE}
\email{S.Kraus@exeter.ac.uk}

\author[0000-0001-9253-7785]{Karolina Kubiak}
\affiliation{\CHARA}
\email{kkubiak@gsu.edu}

\begin{abstract}
Yellow hypergiants are rare, evolved massive stars generally
interpreted as post-red-supergiant objects evolving blueward through
or near the dynamically unstable Yellow Evolutionary Void. V509~Cassiopeiae  is a remarkable
example: its spectroscopically inferred effective temperature
increased from approximately 5000~K in the mid-1980s to about
7900~K by the late 1990s and has since remained near this value
during its post-heating state. We use simultaneous six-telescope CHARA
observations obtained in 2023 with MIRC-X in the $H$ band and MYSTIC
in the $K$ band to measure its photospheric diameter, search for a
luminous companion, and constrain any spatially extended CO
atmosphere.
Interferometric data fitting with fixed atmosphere-model center-to-limb variation profiles
yield limb-darkened angular diameters of
$1.1840$--$1.1950$~mas. Although nominally about 5\% smaller than
the previous Palomar Testbed Interferometer measurement, the difference is only $1.8\sigma$,
indicating no significant secular change over 17--21~yr. The closure
phases are consistent with a centrosymmetric brightness distribution,
and no near-infrared-bright companion is detected over the sampled
separations. The median $3\sigma$ limits are $\Delta H=5.28$~mag and
$\Delta K=6.48$~mag. The B1~V companion proposed from ultraviolet
spectroscopy is expected to lie below these limits and is 
not excluded.
The CO first-overtone channels give a weighted mean
CO-to-continuum diameter ratio of $1.011\pm0.004$, below the adopted
$3\sigma$ detection criterion. We  find no significant
CO-band angular extension. V509~Cas is thus hot and molecularly
compact despite retaining spectroscopically visible atomic
circumstellar gas.
\end{abstract}


\section{Introduction}
\label{sec:introduction}

Yellow hypergiants (YHGs) are short-lived, highly luminous massive
stars generally interpreted as evolving blueward after the
red-supergiant phase
\citep{deJager1998,Nieuwenhuijzen2012,Jones2025}.
Their evolution is complicated by the Yellow Evolutionary Void, an evolutionary phase, in which the outer stellar layers become dynamically
unstable. Rather than
crossing this region smoothly, YHGs can undergo strong pulsations,
enhanced winds, atmospheric expansion and contraction, and episodic
outbursts. These processes may create an optically thick wind or
pseudo-photosphere, causing the apparent effective temperature,
spectral type, and radius to change without requiring equally rapid
evolution of the stellar interior
\citep{deJager1998,Nieuwenhuijzen2012,Kasikov2024}.

V509~Cassiopeiae (HR~8752, HD~217476, Table~\ref{tab:target_properties}) provides one of the
best-studied examples of rapid apparent evolution across the
yellow-hypergiant temperature range. During the mid-1980s it displayed a
late-type spectrum corresponding to an effective temperature of
approximately 5000~K. By the late 1990s its spectroscopically inferred
temperature had reached about 7900~K
\citep{Luck1975,LambertLuck1978,Israelian1999,Nieuwenhuijzen2012,Lobel2013}. Thus, its apparent
temperature increased by nearly 3000~K in little more than a decade.
Such a rapid change is much shorter than the expected evolutionary
timescale of the stellar interior and is  generally
associated with a major restructuring of the extended atmosphere.
A decline in wind density and opacity can move the apparent
photosphere inward to hotter layers, while changes in mass loss,
pulsation, and atmospheric ionization can alter the observed spectrum
\citep{Israelian1999,Nieuwenhuijzen2012,Lobel2013}. V509~Cas has since remained near its hot spectroscopic state, with a
comparatively stable long-term mean brightness but continued
pulsational variability
\citep{Klochkova2019,Kasikov2024}.

A direct angular-diameter measurement provides an independent test of
this interpretation. If the historical increase in temperature
corresponded to a decrease in the apparent photospheric radius, the
star might have continued to contract after reaching the hot state.
The Palomar Testbed Interferometer (PTI) measured
$\theta_{\rm UD}=1.224\pm0.031$~mas and inferred
$\theta_{\rm LD}=1.245\pm0.032$~mas
\citep{vanBelle2009}. Archival PTI records show that these observations
were obtained in 2002 and 2006, after the rapid temperature increase.
A new measurement  does not directly sample the earlier
heating episode, but it tests whether the apparent photospheric
diameter continued to evolve during the subsequent hot state. Our
2023 Center for High Angular Resolution Astronomy (CHARA) Array observations extend this temporal baseline by
17--21~yr.

The unusual spectrum of V509~Cas has also led to suggestions that it
may contain a hot companion. An ultraviolet excess was interpreted as
evidence for a hot secondary by \citet{SticklandHarmer1978}, while
contemporaneous spectroscopy revealed substantial line-profile and
spectral variations that complicated the interpretation of the
system \citep{LambertLuck1978}. Subsequent long-term spectroscopic
monitoring confirmed the complex and time-variable atmospheric
behavior of V509~Cas but did not establish an unambiguous orbital
solution or a secure set of secondary spectral lines
\citep{Percy1992,Lobel2013,Klochkova2019}. A spatially resolved companion search
 provides a complementary test. In particular, a companion
contributing substantially to the optical or near-infrared flux could
affect the inferred spectral energy distribution and complicate the
interpretation of the present $\teff\simeq7900$~K state. 
%A near-infrared non-detection cannot exclude every possible companion, but it can place quantitative limits on luminous companions over the angular separations accessible to CHARA.

A third question concerns the present circumstellar environment.
YHG outbursts can eject dense material that subsequently cools,
forms molecules, and, under suitable conditions, condenses into dust.
The resulting molecular layers can extend well beyond the continuum
photosphere and remain detectable after the photometric event itself.
V509~Cas retains persistent permitted and forbidden emission lines,
demonstrating that atomic circumstellar gas is still present
\citep{Klochkova2019,Kasikov2024}. The long-term stability of the
[O~I] and [Ca~II] emission-line profiles has been interpreted as
evidence for a persistent disk- or ring-like circumstellar structure
\citep{Kasikov2024}. The remaining question is whether this atomic
environment is accompanied by compact continuum emission or a
spatially extended molecular atmosphere.

Near-infrared spectroscopy provides complementary evidence for a
change in the molecular environment of V509~Cas. Variable CO
first-overtone emission and absorption were observed in 1979--1980,
but the CO features had disappeared by 1988 and were not detected in
spectra obtained in 2003, 2004, or 2019
\citep[see review ][]{Kraus2023}. The CHARA observations  test whether
spatially extended structure remains in the CO-band channels despite
the absence of prominent spectroscopic CO features.

Near-infrared spectro-interferometry can separate the continuum
photosphere from wavelength-dependent circumstellar structure. In
cool luminous stars, the apparent angular diameter commonly increases
across the CO first-overtone bands because those channels sample an
extended CO-forming atmosphere
\citep{Perrin2005MuCep,Ohnaka2011Betelgeuse,
Wittkowski2012VYCMa,ArroyoTorres2013,Ohnaka2013Antares,
Gravity2021GCIRS7}. We characterize the molecular extension using
$R_{\rm CO}\equiv\theta_{\rm CO}/\theta_{\rm phot}$, where
$\theta_{\rm phot}$ and $\theta_{\rm CO}$ are the continuum and
CO-channel angular diameters, respectively. Published measurements of cool supergiants commonly show CO-to-continuum diameter ratios of
$R_{\rm CO}\simeq1.2$--$1.7$. A VLTI/AMBER survey reported increasing molecular extension with
luminosity within its cool-supergiant sample, but the behavior of
hotter YHGs remains poorly constrained
\citep{ArroyoTorres2015}.

\begin{deluxetable*}{lccc}
\tablecaption{Adopted Properties of the Three CHARA-resolved
Hypergiants\label{tab:target_properties}}
\tabletypesize{\small}
\tablehead{
\colhead{Property} &
\colhead{V509 Cas (HR 8752)\tablenotemark{a}} &
\colhead{$\rho$ Cas\tablenotemark{b}} &
\colhead{RW Cep\tablenotemark{c}}
}
\startdata
Spectral type
    & A6~Ia$^{+}$
    & F--G~Ia$^{+}$
    & K2--M2~Ia--0 \\
$\teff$ (K)
    & $7900\pm200$
    & $\sim7000$
    & $3900$--$4400$ \\
Distance (kpc)\tablenotemark{d}
    & $3.368\pm0.127$
    & $2.810^{+0.104}_{-0.102}$
    & $3.921^{+0.168}_{-0.157}$ \\
Radius ($R_\odot$)\tablenotemark{e}
    & $433\pm17$
    & $634\pm24$
    & $1105^{+49}_{-46}$ \\
$\log(L/L_\odot)$\tablenotemark{f}
    & $5.83^{+0.05}_{-0.06}$
    & $5.83^{+0.12}_{-0.16}$
    & $5.36^{+0.10}_{-0.13}$ \\
$\langle\theta_{\rm UD}\rangle_{\rm cont}$ (mas)\tablenotemark{g}
    & $1.164\pm0.002$
    & $2.10^{+0.01}_{-0.02}$
    & $2.62\pm0.03$ \\
$\rco$\tablenotemark{h}
    & $1.011\pm0.004$
    & $1.39\pm0.09$
    & $1.66\pm0.05$ \\
PTI $\theta_{\rm LD}$ (mas)
    & $1.245\pm0.032$
    & \nodata
    & \nodata \\
Present state
    & \shortstack{Hot state; atomic\\line-emitting gas}
    & \shortstack{Recurrently eruptive;\\extended CO atmosphere}
    & \shortstack{Great Dimming and recovery;\\dust and extended CO atmosphere} \\
\enddata
\tablenotetext{a}{
For V509~Cas, the spectral classification and $T_{\rm eff}$ are
adopted from \citet{Kasikov2024}, and the PTI diameter from
\citet{vanBelle2009}. The CHARA angular-diameter and CO-extension
results are from this work.
}
\tablenotetext{b}{
For $\rho$~Cas, the spectral classification and approximate
warm-state $T_{\rm eff}$ are adopted from
\citet{vanGenderen2025}. The CHARA angular diameter and
$\rco$ are from \citet{Anugu2024RhoCas}.
}
\tablenotetext{c}{
For RW~Cep, the spectral classification, temperature range,
CHARA angular diameter, and $\rco$ are adopted
from \citet{Anugu2024RWCep}.
}
\tablenotetext{d}{
Group-based distances are adopted from \citet{Kasikov2026}.
They were determined from Gaia parallaxes of nearby co-moving stars
associated with the environments of the three targets.
}
\tablenotetext{e}{
Radii were calculated in this work from $R=\theta d/2$, using the
group-based distances of \citet{Kasikov2026}. For V509~Cas, we used
the adopted SATLAS fixed-CLV diameter from
Table~\ref{tab:photosphere_fits}. For $\rho$~Cas and RW~Cep, we used
the CHARA photospheric diameters from
\citet{Anugu2024RhoCas} and \citet{Anugu2024RWCep}, respectively.
The uncertainties propagate the quoted distance and angular-diameter
uncertainties.
}
\tablenotetext{f}{
Luminosities are adopted from \citet{Kasikov2026}
and are not recalculated from the CHARA-based radii listed
here. Their derivation combines distances with literature
angular diameters and effective temperatures obtained at
different epochs. Because yellow hypergiants are variable,
the quoted uncertainties may not include all systematic
effects associated with pulsation phase, temperature, and
angular-diameter prescription. In particular, the published
RW~Cep luminosity uses the JSDC2 photometric angular
diameter rather than the CHARA measurement.
}
\tablenotetext{g}{
For V509~Cas, the listed value is the weighted mean uniform-disk
diameter of the channel-binned $K$-band continuum measurements used to
calculate the differential CO-to-continuum ratio. The global $K$-band
uniform-disk fit and the adopted SATLAS photospheric diameter are
given in Table~\ref{tab:photosphere_fits}.
}
\tablenotetext{h}{
For V509~Cas, the listed value is the measured effective,
channel-integrated CO-to-continuum diameter ratio. For comparison with
published molecular extensions, $\rco=1.05$ is adopted as a conservative
sensitivity threshold; it is not a formal confidence or completeness limit.
The $\rho$~Cas and RW~Cep values are direct CHARA measurements from
\citet{Anugu2024RhoCas} and \citet{Anugu2024RWCep}, respectively.
}
\tablecomments{
The table is restricted to the three yellow or cool hypergiants
resolved with CHARA and is not intended as a census of Galactic
hypergiants. Spectral classifications and several fundamental
parameters are phase dependent and should be regarded as
representative values.
}
\end{deluxetable*}

In this work, we present simultaneous six-telescope CHARA/MIRC-X
$H$-band and MYSTIC $K$-band observations of V509~Cas obtained in
2023. We use these data to test whether the photospheric diameter continued
to evolve after the rapid heating episode, to search for a luminous
resolved companion, and to measure any CO-band angular enlargement
relative to the adjacent continuum. We then compare V509~Cas with
$\rho$~Cas and RW~Cep to place its present hot, comparatively stable state in the
broader context of YHG atmospheric evolution and episodic mass loss.

The adopted stellar properties and the principal interferometric
measurements for V509~Cas, $\rho$~Cas, and RW~Cep are summarized in
Table~\ref{tab:target_properties}.

The remainder of this paper is organized as follows.
Section~\ref{sec:observations} describes the CHARA observations and data
reduction, and Section~\ref{sec:methods} presents the continuum-diameter,
CO-band extension, and companion-search analyses.
Section~\ref{sec:results} reports the resulting photospheric diameters,
molecular-atmosphere constraints, and companion-detection limits.
Section~\ref{sec:MolecularAtmospheresComparison} compares the CO-band
angular extension of V509~Cas with published measurements of other
luminous evolved stars.
Section~\ref{sec:discussion} discusses the implications for the present
hot state of V509~Cas and its relation to other YHGs, and
Section~\ref{sec:conclusions} summarizes our main conclusions.

%%%%%%%%%%%%%%%%%%%%%%%%%%%%%%%%%%%%%%%%
\section{Observations and Data Reduction}
\label{sec:observations}
The CHARA Array is a
long-baseline optical/near-infrared interferometer located on Mount
Wilson, California. It consists of six 1-m telescopes distributed in a
Y-shaped configuration, providing baselines up to 331~m
\citep{tenBrummelaar2005}. At the longest baselines, the corresponding
nominal angular resolution is approximately 0.5~mas in the $H$ band and
0.7~mas in the $K$ band.

V509~Cas was observed with all six CHARA telescopes on UT 2023 October
28. MIRC-X provided spectrally dispersed $H$-band fringes from 1.47 to
1.80~$\mu$m at a resolving power of $R\simeq190$, corresponding to
32 spectral channels \citep{Anugu2020MIRCX}. MYSTIC simultaneously
recorded $K$-band fringes from 2.00 to 2.37~$\mu$m at
$R\simeq278$, corresponding to 62 spectral channels
\citep{Monnier2018MYSTIC,Setterholm2023}. The six-telescope configuration delivered
15 squared visibilities and 20 closure phases per spectral channel,
providing simultaneous constraints on the wavelength-dependent source
size and departures from centrosymmetry.

The data were reduced with the standard MIRC-X/MYSTIC reduction
pipeline \citep{Anugu2020MIRCX} and calibrated using interleaved observations
of 7~And, HD~13137, and HD~554. Uniform-disk calibrator diameters were
adopted from the JMMC Stellar Diameters Catalogue
\citep{Chelli2016,Bourges2017}. The observing configuration and calibrator
properties are summarized in Table~\ref{tab:observations}.

The raw data are available through the CHARA Observers Database\footnote{\url{https://chara.gsu.edu/observers/database}}. For
reproducibility, the calibrated OIFITS files, fitting scripts, and
figures used in this work have been deposited in Zenodo and will become
public upon publication \citep{anugu_2026_Zenodo_21811660}.

\begin{figure*}[t]
\centering
\digitalasset
\includegraphics[width=\textwidth]{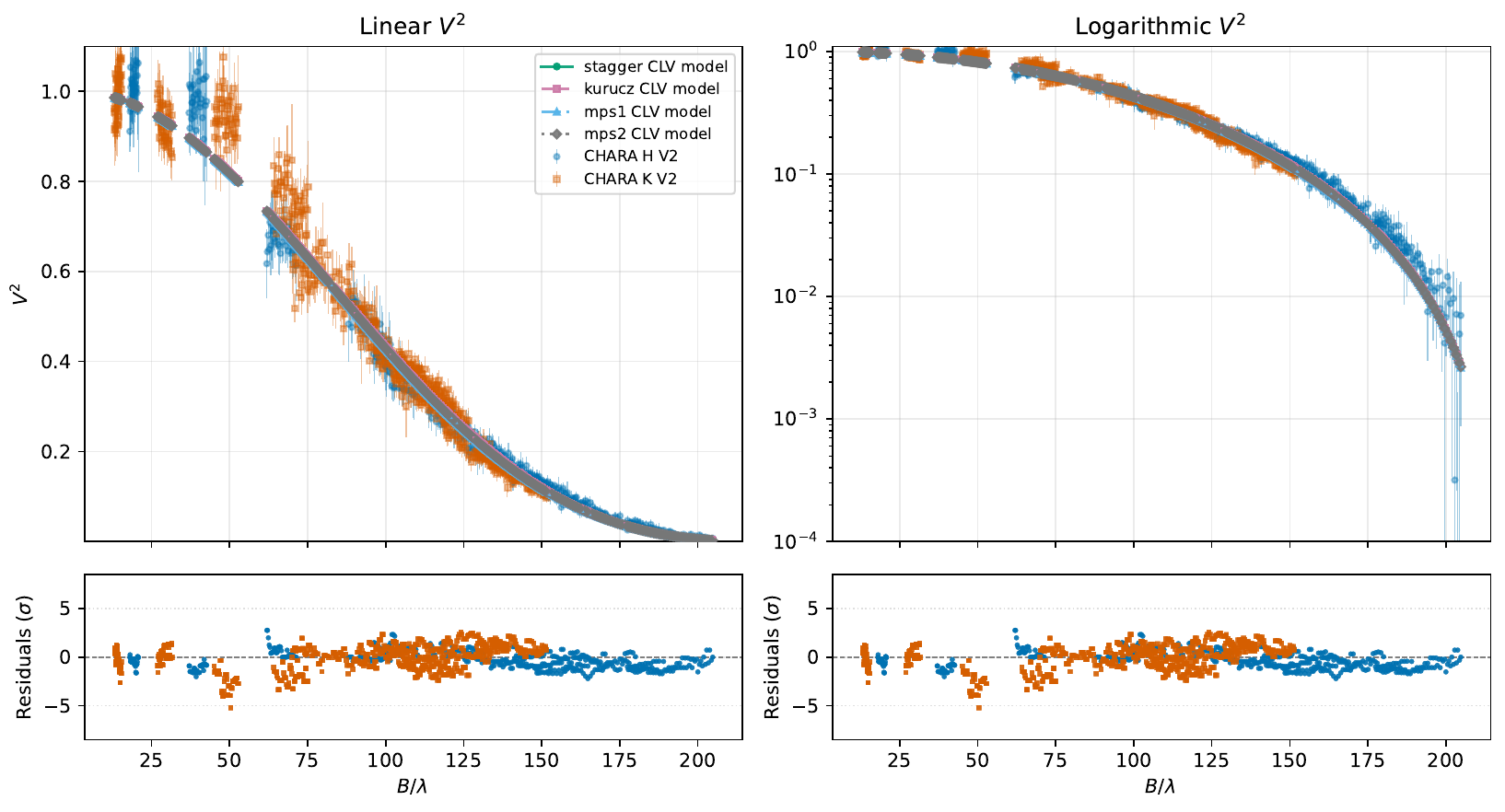}
\caption{
Squared visibilities and normalized residuals for V509~Cas as a
function of spatial frequency. Blue points show the MIRC-X
$H$-band measurements, and orange points show the MYSTIC $K$-band
measurements. The colored curves represent the fixed-CLV
photospheric models listed in Table~\ref{tab:photosphere_fits}.
The complete Figure Set (2 images), showing the band-specific analytic
limb-darkening fits and residuals, is available in the online journal
and in the Zenodo archive.
The larger uncertainties in some $V^2$ measurements are associated
with delay-line tracking errors caused by timing jitter
\citep{Anugu2026JATIS}.
}\label{fig:visibility}
\end{figure*}

\begin{deluxetable*}{lcccc}
\tablecaption{Summary of the CHARA Observations and Interferometric
Calibrators\label{tab:observations}}
\tabletypesize{\small}
\tablehead{
\colhead{UT Date} &
\colhead{Instrument} &
\colhead{Band} &
\colhead{Wavelength Range} &
\colhead{Spectral Resolution}
}
\startdata
\multicolumn{5}{c}{\textit{Observing configuration}} \\
\hline
2023 Oct 28 & MIRC-X & $H$ & 1.47--1.80~$\mu$m & $R\simeq190$ \\
2023 Oct 28 & MYSTIC & $K$ & 2.00--2.37~$\mu$m & $R\simeq278$ \\
\hline
\multicolumn{5}{c}{\textit{Interferometric calibrators}} \\
\multicolumn{1}{c}{Calibrator} &
\multicolumn{1}{c}{Separation} &
\multicolumn{1}{c}{$\theta_{\rm UD}$ (mas)} &
\multicolumn{1}{c}{$\sigma_\theta$ (mas)} &
\multicolumn{1}{c}{Catalog} \\
\hline
7 And    & $9.6^\circ$ & 0.69267 & 0.06762 & JSDC2 \\
HD 13137 & $7.0^\circ$ & 0.66210 & 0.04627 & JSDC2 \\
HD 554   & $4.3^\circ$ & 0.58147 & 0.01290 & JSDC2 \\
\enddata
\tablecomments{
All six CHARA telescopes, S1--S2--E1--E2--W1--W2, were used.
Calibrator separations are angular distances from V509~Cas.
The calibrator angular diameters are uniform-disk estimates from
JSDC2 \citep{Bourges2017}.
}
\end{deluxetable*}

%%%%%%%%%%%%%%%%%%%%%%%%%%%%%%%%%%%%%%%%%
\section{Interferometric Modeling}
\label{sec:methods}

\subsection{Continuum photosphere}
\label{sec:methods_photosphere}

We modeled the calibrated squared visibilities ($V^2$)  with \texttt{PMOIRED}\footnote{https://github.com/amerand/PMOIRED.git} \citep{Merand2022PMOIRED}, a widely used   Python package for parametric interferometric model fitting. 
Our limb-darkening fitting approach was inspired by \citet{Anugu2026LimbDarkening}, and the implementation was adapted from the accompanying publicly released code \citep{anugu_2026_Zenodo}. The photospheric analysis uses the retained MIRC-X channels from
1.50 to 1.75~$\mu$m and the MYSTIC continuum channels from 2.00 to 2.30~$\mu$m. Circular uniform disks were first fitted independently in $H$ and $K$.

We then performed a joint $H+K$ power-law fit using
\begin{equation}
\frac{I(\mu)}{I(1)}=\mu^\alpha,
\end{equation}
where $\mu=\cos\gamma$ and $\gamma$ is the angle between the local
surface normal and the line of sight. Independent limb-darkening
coefficients, $\alpha_H$ and $\alpha_K$, were used for the two bands.
Both coefficients were treated as free
parameters with atmosphere-informed priors derived from the predictions
of the Kurucz ATLAS \citep{Kurucz1993}, Merged Parallelised
Simplified-ATLAS set~1 and set~2 (MPS1 and MPS2;
\citealt{Kostogryz2022}), Stagger
\citep{Magic2013,Magic2015}, and spherical SATLAS
\citep{Neilson2013} atmosphere grids. We adopted stellar input
distributions of $T_{\rm eff}=7900\pm100$~K, $\log g=0.34\pm0.20$
(cgs), and $[\mathrm{Fe}/\mathrm{H}]=0.00\pm0.20$ to calculate the
bandpass-integrated coefficients with the ExoTiC-LD package\footnote{https://github.com/nanugu/ExoTiC-LD.git}
\citep{Grant2024JOSS}.

The remaining free parameters were a common angular diameter,
$\theta_{\rm PL}^{H+K}$, and a multiplicative squared-visibility
normalization, $V_0^2$, such that
\begin{equation}
V_{\rm fit}^2 = V_0^2 V_{\rm PL}^2,
\qquad
V_{\rm PL}^2(B=0)=1.
\end{equation}
The parameter $V_0^2$ is a single common multiplicative normalization
applied to all fitted baselines and both bands; it accounts for any residual
offset in the calibrated visibility scale and has an ideal value of unity. Because the data
provide limited higher-lobe leverage, the fitted limb-darkening
coefficients remain correlated with angular diameter and are 
used primarily as diagnostic measurements.

We then fitted the visibilities directly with fixed,
bandpass-integrated center-to-limb variation (CLV) profiles,
$I(\mu)$, from the Kurucz, MPS1, MPS2, Stagger, and SATLAS atmosphere
grids. For each grid $g$, the CLV shape was held fixed, while only the
angular diameter, $\theta_{\rm CLV}^{g}$, and the common visibility normalization,
$V_0^2$, were varied. The spherical SATLAS angular scale was converted
to its corresponding Rosseland angular diameter. We adopt the
SATLAS-derived Rosseland result
as the adopted reference photospheric diameter and use the spread among the grid
results to characterize the model dependence. The Kurucz result is
retained for the most nearly like-for-like comparison with the
Kurucz-corrected PTI diameter.

Closure phases were fitted independently, rather than included in the $V^2$ photospheric fits, to test for departures from point symmetry (see Section~\ref{sec:methods_companion}).

%%%%%%%%%%%%%%%%%%%%%%%%%%%%%%%%%%%%%%%%%%%%%%%%%
\subsection{CO-band angular-diameter extension}
\label{sec:methods_co}

We divided the MYSTIC data into continuum channels from 2.00 to 2.30~$\mu$m and CO first-overtone channels from 2.30 to 2.37~$\mu$m. Each subset was fitted independently  to obtain the effective continuum and CO-channel diameters, $\theta_{\rm phot}$ and $\theta_{\rm CO}$. We define
\begin{equation}
\mathcal{R}_{\rm CO}=\frac{\theta_{\rm CO}}{\theta_{\rm phot}}.
\end{equation}
This differential measurement tests whether the CO channels are measurably larger than the adjacent continuum; it does not require that all CO-bearing gas be represented by a single physical shell.

To measure diameter as function of wavelength, we
also fitted pairs of adjacent spectral channels independently. We
first fitted a uniform-disk diameter in each two-channel bin. We then
repeated the analysis with fixed power-law limb-darkening coefficients,
$\alpha_H=0.128$ and $\alpha_K=0.106$, adopted from the
bandpass-integrated Kurucz profiles. In both cases, only the angular
diameter was varied within each spectral bin. The fixed-limb-darkening
fits alter the absolute angular scale but preserve the differential
wavelength dependence.

We evaluated the sensitivity using injection--recovery simulations.
Synthetic observables were generated with \texttt{PMOIRED} the measured
$(u,v,\lambda)$ sampling and the same noise characteristics, while
the input CO-to-continuum diameter ratio was varied from
$R_{\rm CO}=1.00$ to 1.50 in steps of 0.05. Each simulated data set
was refitted using the same two-subset diameter procedure applied to
the observations. All injected models with
$R_{\rm CO}\geq1.05$ were recovered in this test.

\subsection{Companion search and detection limits}
\label{sec:methods_companion}

We performed a two-dimensional companion search with
\texttt{PMOIRED}, fitting the squared visibilities and closure phases
jointly. The search sampled a Cartesian grid with a spacing of
0.5~mas over projected separations out to 100~mas. This search radius
was chosen to remain within the approximate spectral-coherence field,
$\lambda\mathcal{R}/B \simeq 180$--$270$~mas, and thereby limit
bandwidth-smearing losses. At each grid location, the companion flux
ratio was optimized simultaneously with the photospheric parameters.
Because bandwidth smearing and the coherent field vary with wavelength
and angular separation, the sensitivity is not uniform across the
search region. We then derived the contrast limits from
injection--recovery.

\begin{figure*}[t]
\centering
\includegraphics[width=\linewidth]{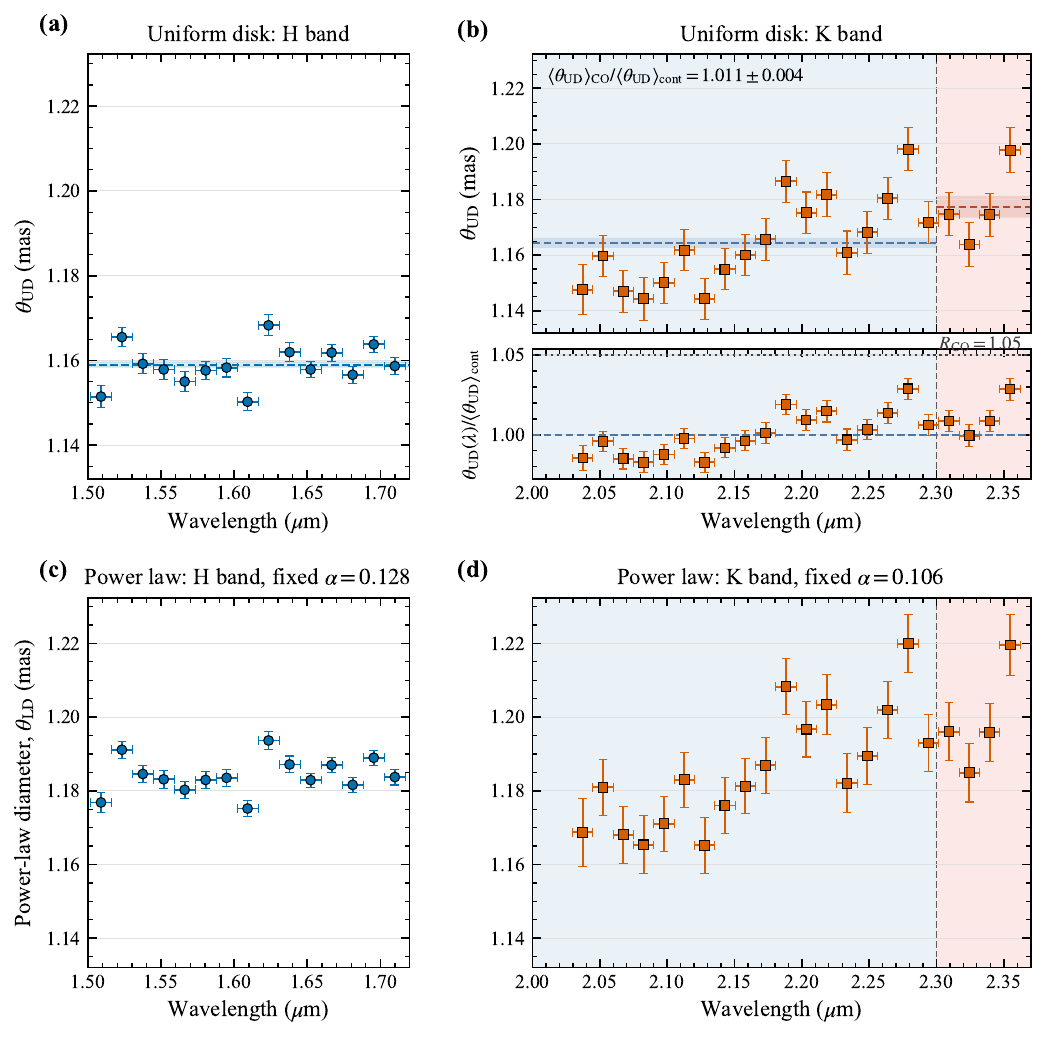}
\caption{
Effective angular diameter of V509~Cas as a function of wavelength,
obtained by fitting pairs of adjacent spectral channels.
Panels~(a) and (b) show uniform-disk fits to the MIRC-X $H$-band and
MYSTIC $K$-band data, respectively. In panel~(b), the blue and red
shading mark the $K$-band continuum and CO first-overtone regions, and
the dashed horizontal lines indicate their weighted mean diameters.
The weighted mean ratio is
$\langle\theta_{\rm UD}\rangle_{\rm CO}/
 \langle\theta_{\rm UD}\rangle_{\rm cont}
=1.011\pm0.004$.
The lower subpanel of panel~(b) shows each $K$-band diameter normalized
by the continuum mean; the dotted line at 1.05 marks the adopted
conservative sensitivity threshold.
Panels~(c) and (d) show the corresponding $H$- and $K$-band fits using Kurucz grid
fixed power-law limb-darkening coefficients of
$\alpha_H=0.128$ and $\alpha_K=0.106$, respectively.
The limb-darkened fits change the absolute angular scale but preserve
the same wavelength dependence. No significant diameter increase is
detected across the CO channels.
}
\label{fig:diameter_wavelength}
\end{figure*}

\begin{deluxetable*}{llll}
\tablecaption{Photospheric Angular-diameter Fits\label{tab:photosphere_fits}}
\tabletypesize{\scriptsize}
\tablehead{
\colhead{Fit or atmosphere} &
\colhead{$\theta$ (mas)} &
\colhead{Limb-darkening treatment} &
\colhead{Role}
}
\startdata
Uniform disk, $H$ & $1.148\pm0.003$ & No limb darkening & Single-band diagnostic \\
Uniform disk, $K$ & $1.173\pm0.007$ & No limb darkening & Single-band diagnostic \\
Joint power law & $1.181\pm0.006$ & Free $\alpha_H$ and $\alpha_K$ & Atmosphere-prior fit \\
Kurucz CLV & $1.1862\pm0.0036$ & Fixed tabulated $H+K$ CLVs & Used to compare with PTI \\
Stagger CLV & $1.1850\pm0.0036$ & Fixed tabulated $H+K$ CLVs & Model comparison \\
MPS1 CLV & $1.1840\pm0.0037$ & Fixed tabulated $H+K$ CLVs & Model comparison \\
MPS2 CLV & $1.1843\pm0.0054$ & Fixed tabulated $H+K$ CLVs & Model comparison \\
SATLAS CLV & $1.195\pm0.007$ & Fixed tabulated $H+K$ CLVs & Adopted reference photospheric scale \\
\enddata
\tablecomments{For the joint power-law fit, $\theta$, $\alpha_H$, $\alpha_K$, and $V_0^2$ were free. Gaussian priors on $\alpha_H$ and $\alpha_K$ were centered on the corresponding Kurucz limb darkening predictions, with widths defined in Section~\ref{sec:methods_photosphere}. For the fixed-CLV fits, the tabulated bandpass intensity profiles were held fixed and only $\theta_{\rm CLV}^{g}$ and $V_0^2$ were varied. $V_0^2=1.01$ is a visibility-scale parameter and is omitted from the table.}
\end{deluxetable*}

\begin{deluxetable}{lcccc}
\tablecaption{Companion Detection Limits}
\tablehead{
\colhead{Band} &
\colhead{$\Delta m_{50\%}$} &
\colhead{$f_{50\%}$} &
\colhead{$\Delta m_{90\%}$} &
\colhead{$f_{90\%}$}
}
\startdata
$H$ & 5.28 & 0.78\% & 5.06 & 0.95\% \\
$K$ & 6.48 & 0.26\% & 6.27 & 0.31\% \\
\enddata
\label{tab:companion_limits}
\tablecomments{
Magnitude differences correspond to $3\sigma$
injection--recovery thresholds. The 50\% and 90\% columns give the
contrasts brighter than which 50\% and 90\% of injected companions
were recovered, respectively.
}
\end{deluxetable}

%%%%%%%%%%%%%%%%%%%
\section{Results}
\label{sec:results}

\subsection{Photospheric angular diameter}
\label{sec:results_photosphere}

The continuum photosphere is clearly resolved in both bands (Figure~\ref{fig:visibility}). Independent uniform-disk fits give $\theta_{\rm UD,H}=1.148\pm0.003~\mas$ and
$\theta_{\rm UD,K}=1.173\pm0.007~\mas$.

The $K$-band value is 2.2\% larger than the $H$-band value, showing  stronger limb darkening in H-band than K-band. The joint power-law fit gives $\theta_{\rm PL}^{H+K}=1.181\pm0.006$~mas, $\alpha_H=0.180\pm0.039$, and $\alpha_K=0.025\pm0.029$. The limb-darkening coefficients remain weakly constrained as $V^2$ do not extend over first-null.

The fixed-CLV fits yield joint $H+K$ angular diameters from 1.1840 to 1.1950~mas for different atmospheric model grids (Table~\ref{tab:photosphere_fits}). We adopt the SATLAS value,
$\theta_{\rm CLV}^{H+K}=1.195\pm0.007$~mas, as the reference
photospheric diameter. We use the Kurucz value,
$1.1862\pm0.0036$~mas, for the historical comparison because the
PTI limb-darkening correction was also based on Kurucz atmosphere
models. The quoted statistical uncertainties were estimated through
bootstrap resampling.

%%%%%%%%%%%%%%%%%%%%%%%%%%%%%%%%%%%%%
\subsection{CO-band angular-extension limit}
\label{sec:results_co}

Figure~\ref{fig:diameter_wavelength} shows the effective
uniform-disk diameter obtained from pairs of adjacent spectral
channels. The weighted mean CO-to-continuum diameter ratio is
\begin{equation}
\frac{\langle\theta_{\rm UD}\rangle_{\rm CO}}
     {\langle\theta_{\rm UD}\rangle_{\rm cont}}
=1.011\pm0.004.
\end{equation}
This corresponds to a marginal increase below the adopted
$3\sigma$ detection criterion, and we  do not claim a
significant CO-band angular extension. Repeating the channel-binned
analysis with fixed power-law limb-darkening coefficients changes the
absolute diameter scale but produces the same wavelength dependence
and no significant discontinuity at 2.30~$\mu$m.

The injection tests recovered all models with
$R_{\rm CO}\geq1.05$. We adopt 1.05 as a conservative
comparison threshold rather than as a formal confidence or
completeness limit. This threshold is well below the extensions
measured for hypergiants $\rho$~Cas  and RW~Cep,
$1.39\pm0.09$ \citep{Anugu2024RhoCas} and $1.66\pm0.05$ \citep{Anugu2024RWCep}, respectively. 

\subsection{Symmetry and continuum emission}
\label{sec:results_symmetry}

\begin{figure*}[t]
\centering
\includegraphics[width=\linewidth]{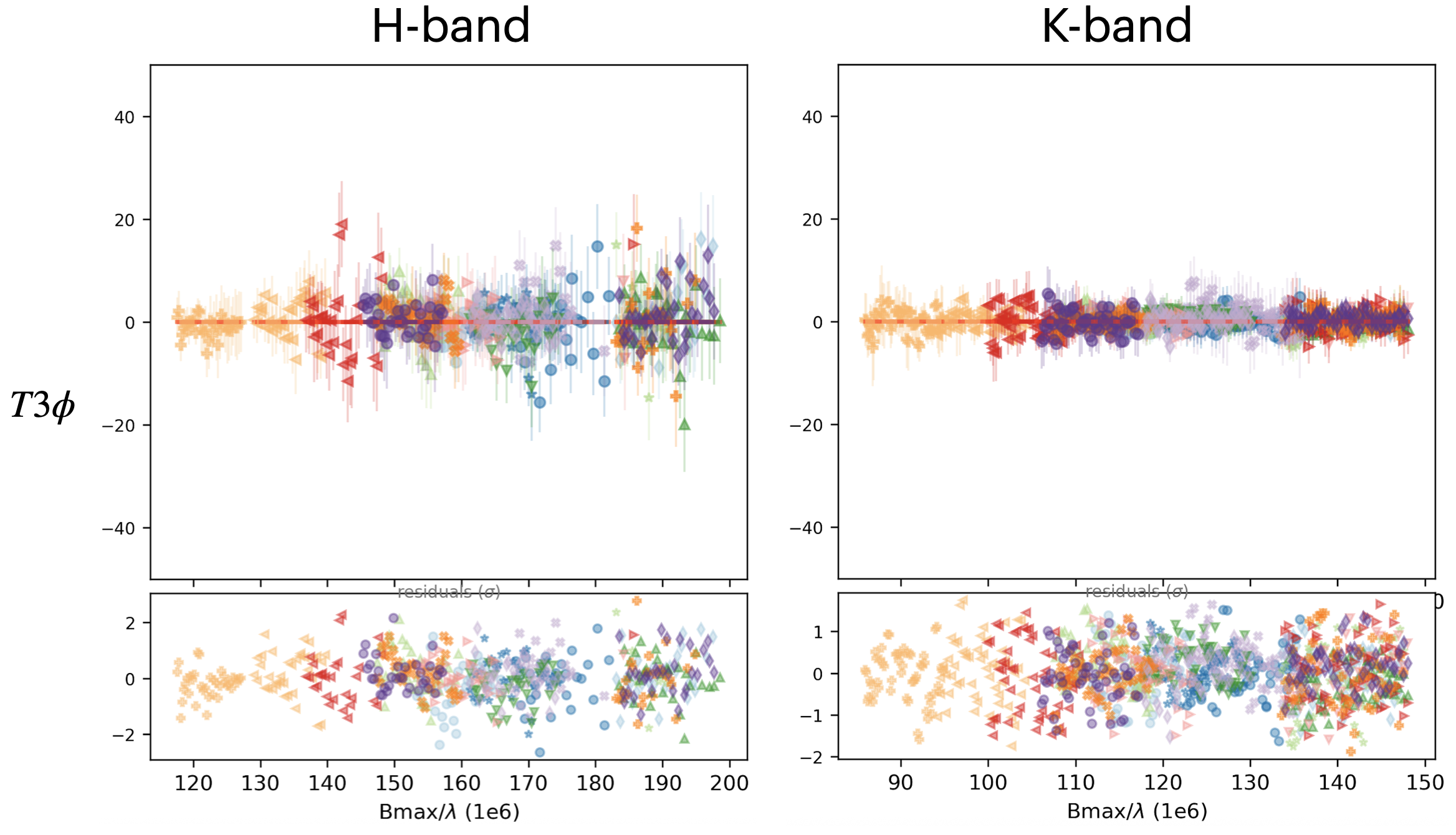}
\caption{
MIRC-X and MYSTIC closure phases and normalized fitting residuals.
The closure phases are consistent with a centrosymmetric
near-infrared brightness distribution. Colors indicate different
telescope triangles.
} \label{fig:closure_phase}
\end{figure*}

The closure phases remain centered near zero in both bands and show no coherent non-centrosymmetric signal (Figure~\ref{fig:closure_phase}). Adding either a fully resolved halo or a circular Gaussian envelope does not yield a statistically compelling improvement over the photosphere-only model. These constraints apply to compact broadband near-infrared emission and do not exclude low-density atomic line emission or colder dust at substantially larger angular scales.

%%%%%%%%%%%%%%%%%%%%%%%%%%%%%%%%%%%
\subsection{Companion limits}
\label{sec:results_companion}

\begin{figure*}[t]
\centering
\includegraphics[width=\textwidth]{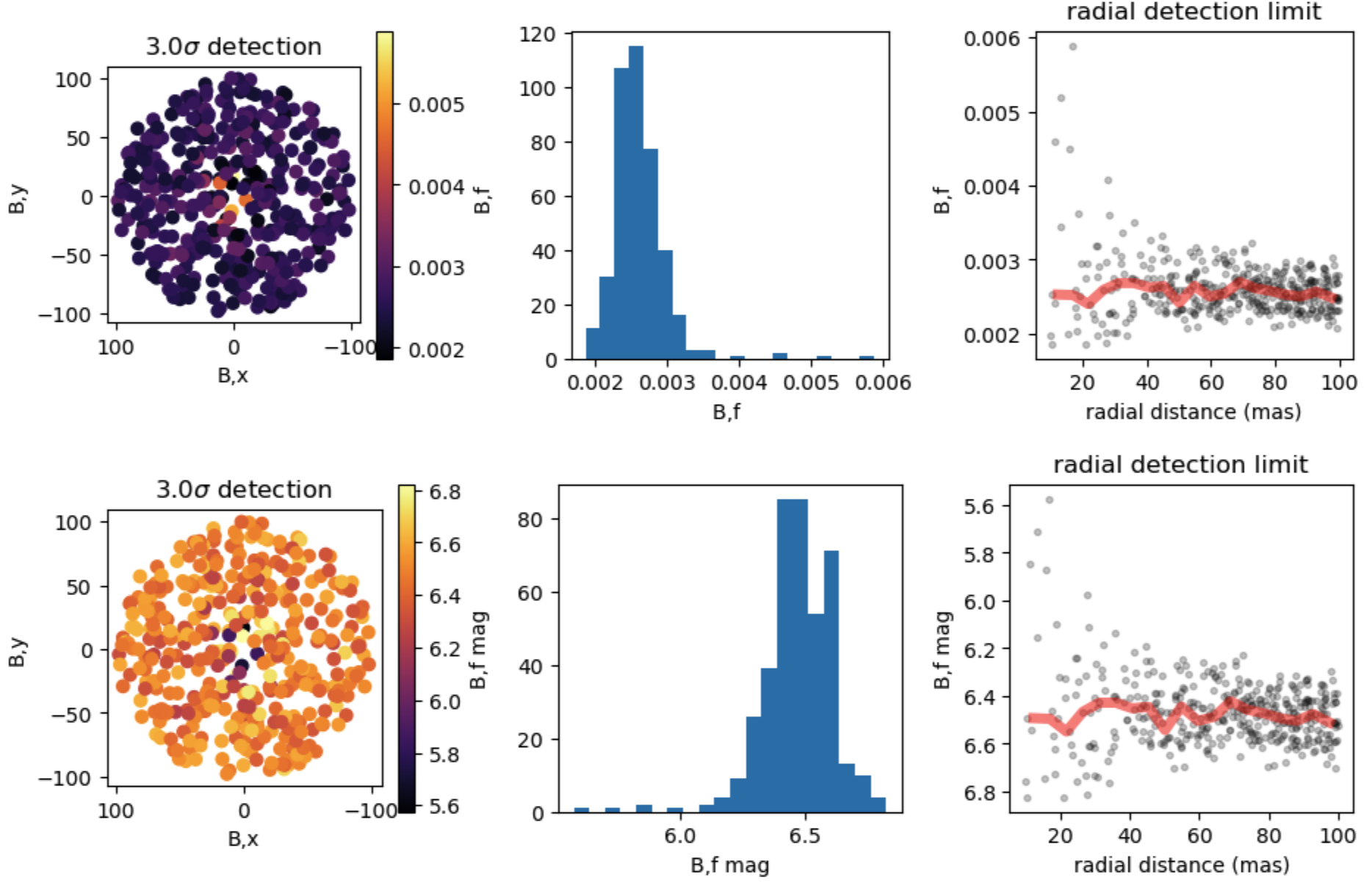}
\caption{
MYSTIC $K$-band companion-search significance map and
injection--recovery detection limits. The top three panels show companion flux ratios, whereas the bottom three panels show the corresponding $\Delta K$-band magnitudes. The median $3\sigma$ threshold
is $\Delta K=6.48$~mag, and 90\% of injected companions were
recovered when brighter than $\Delta K=6.27$~mag. The corresponding
MIRC-X $H$-band limits are listed in Table~\ref{tab:companion_limits}.
The PMOIRED labels $B,x$ and $B,y$ denote the companion's fitted
$x$ and $y$ position offsets, respectively; $B,f$ is its flux ratio
relative to the primary, and $B,f$ mag is the corresponding magnitude
contrast.
}\label{fig:companion_search}
\end{figure*}

No significant companion is detected independently in either the
MIRC-X $H$-band or MYSTIC $K$-band data (Figure~\ref{fig:companion_search} and Table~\ref{tab:companion_limits}). The median $3\sigma$
injection--recovery limits are $\Delta H=5.28$~mag and
$\Delta K=6.48$~mag, corresponding to companion-to-primary flux
ratios of 0.78\% and 0.26\%, respectively. At 90\% recovery
completeness, the corresponding limits are
$\Delta H=5.06$~mag and $\Delta K=6.27$~mag, equivalent to flux
ratios of 0.95\% and 0.31\%, respectively. Because the flux ratio
depends on wavelength and companion spectral type, we retain
separate $H$- and $K$-band limits rather than adopting a single
color-independent contrast.

\section{Comparison with Interferometrically Resolved Molecular Atmospheres}
\label{sec:MolecularAtmospheresComparison}

We compiled published $K$-band continuum and CO angular scales for giant, supergiant, and hypergiant stars, measured with VLTI/AMBER, VLTI/GRAVITY, IOTA/FLUOR, and CHARA/MYSTIC. The literature is heterogeneous: some studies report shell radii, others wavelength-dependent uniform-disk diameters, and others maximum CO-band extensions. The ratio $\rco$ removes distance scaling but not differences in radial model, spectral resolution, or target temporal state.

\begin{figure*}[t]
\centering
\includegraphics[width=\linewidth]{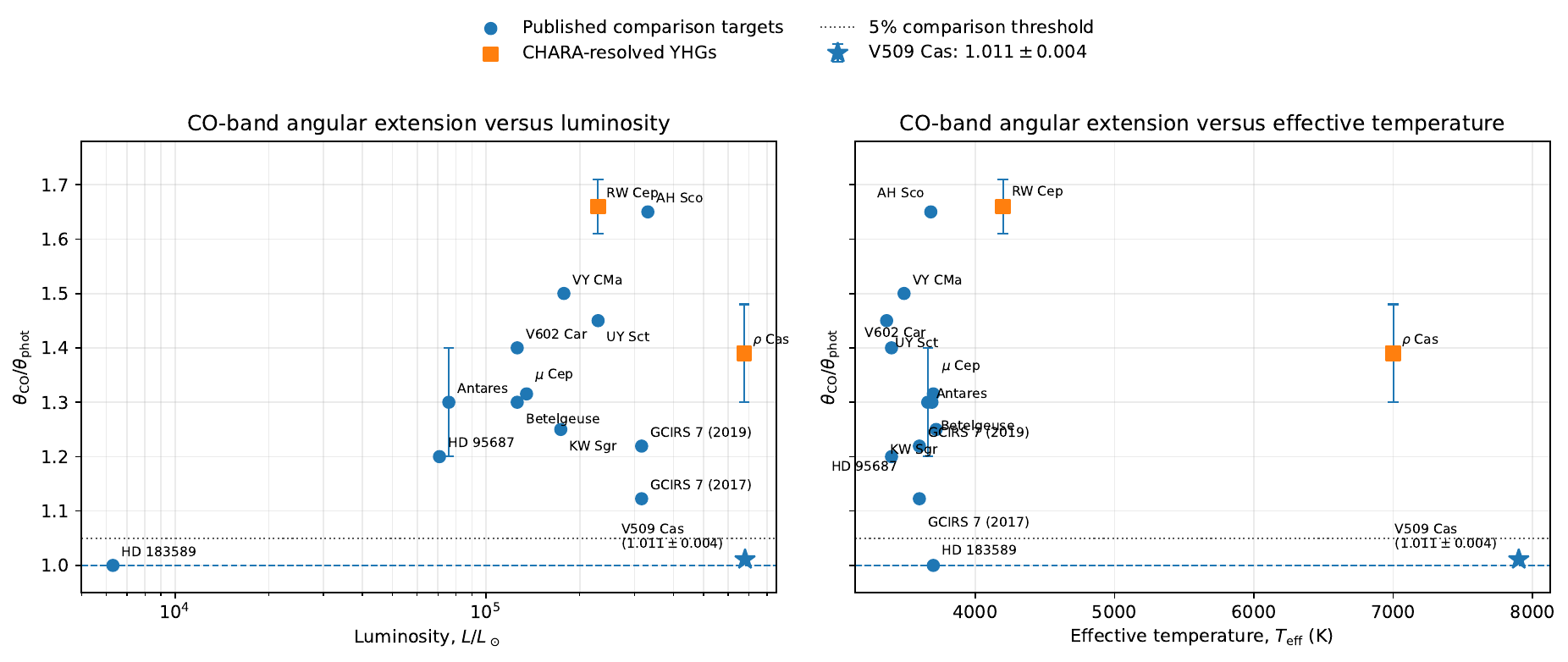}
\caption{Published molecular-atmosphere extension ratios, $\theta_{\rm CO}/\theta_{\rm phot}$, as functions of \textit{(a)} luminosity and \textit{(b)} effective temperature. The compilation includes giant, supergiant, and hypergiant stars observed with VLTI/AMBER, VLTI/GRAVITY, IOTA/FLUOR, and CHARA/MYSTIC. V509~Cas is plotted at its measured value, $\mathcal{R}_{\rm CO}=1.011\pm0.004$, with $L=6.8\times10^5\,L_\odot$ and $\teff=7900$~K; the dotted line marks the adopted 5\% comparison threshold, not a formal confidence or completeness limit. The compilation combines different instruments, epochs, bandpasses, and radial models and is intended as a comparative data set rather than a homogeneous regression sample. Data sources include \citet{Perrin2005MuCep,Ohnaka2011Betelgeuse,Wittkowski2012VYCMa,ArroyoTorres2013,Ohnaka2013Antares,ArroyoTorres2015,Gravity2021GCIRS7,Anugu2024RhoCas,Anugu2024RWCep}.\label{fig:ratio_comparison}}
\end{figure*}

Figure~\ref{fig:ratio_comparison} shows the ratios against luminosity and effective temperature. V509~Cas is plotted at its measured value, $\mathcal{R}_{\rm CO}=1.011\pm0.004$; the figure also marks the adopted 5\% comparison threshold, which is not a formal confidence or completeness limit. The two CHARA comparison hypergiants occupy distinctly extended
states: $\rho$~Cas has $\mathcal{R}_{\rm CO}=1.39\pm0.09$, and RW~Cep has $1.66\pm0.05$.

\section{Discussion}
\label{sec:discussion}

\subsection{A hot, continuum-compact yellow hypergiant}
\label{sec:discussion_state}

The CHARA measurements do not imply an absence of circumstellar
material. Persistent permitted and forbidden emission lines
demonstrate that atomic gas remains around V509~Cas
\citep{Klochkova2019,Kasikov2024}. Instead, the 2023 near-infrared
brightness distribution is dominated by a compact, approximately
centrosymmetric photosphere, with no strong compact continuum halo and no CO-channel angular
enlargement reaching the adopted 5\% comparison threshold. Atomic
line-emitting gas can persist while the near-infrared
continuum and warm molecular-forming regions remain close to the
photospheric radius.

V509~Cas is not unique among yellow hypergiants in lacking reported
near-infrared molecular features. Although YHGs are commonly
interpreted as post-red-supergiant objects, large-scale dusty and/or
cold molecular envelopes have been reported for only about half of
the known population \citep[and references therein]{Kraus2023}. Envelope-bearing Galactic
objects include IRC~+10420, HR~5171A, HD~179821, and Hen~3-1379,
together with several YHGs or candidates in the Magellanic Clouds
and M33 \citep[and references therein]{Kraus2023}. In other systems, material released during relatively brief
mass-loss episodes may expand and dilute without producing a readily
detectable large-scale envelope.

The spatial scales probed by CHARA are distinct from those of the
extended dusty and cold molecular envelopes considered in such
surveys. Our observations constrain the compact, warm CO-forming
region close to the star and do not exclude older or colder ejecta at
substantially larger radii. Infrared observations provide
additional context. Broadband photometry extending from 2.3 to
23~$\mu$m included V509~Cas and showed no substantial mid-infrared
excess \citep{Hackwell1974}. High-resolution spectroscopy of
the 2.3~$\mu$m CO first-overtone region nevertheless revealed strong
circumstellar CO, leading \citet{Lambert1981} to describe
the circumstellar environment as gas-rich but containing little dust.
Optical HST/WFPC2 imaging in 0.373, 0.437, 0.501, and
0.656~$\mu$m,  also revealed no extended circumstellar
nebulosity around V509~Cas \citep{Schuster2006}. Thus, the lack of a
strong infrared excess does not by itself imply an absence of warm
molecular gas: V509~Cas displayed circumstellar CO in the
2.3~$\mu$m first-overtone region during 1979--1980 despite its weak
infrared excess \citep{Lambert1981}.

The CO features subsequently disappeared and were absent in spectra
obtained in 1988, 2003, 2004, and 2019
\citep[and references therein]{Kraus2023}. The 2023 CHARA result
provides a complementary spatial constraint: the CO-band brightness
distribution is not measurably more extended than the adjacent
continuum. The historical absence of a substantial dusty mid-infrared excess
and the present absence of a resolved warm CO atmosphere provide
complementary constraints on different components of the
circumstellar environment. The spectroscopic and interferometric
measurements are both consistent with the absence of a large, warm
molecular atmosphere in the present post-heating state.

%The spatial scales probed by CHARA are distinct from those of the
%extended dusty and cold molecular envelopes considered in such
%surveys. Our observations constrain the compact, warm CO-forming
%region close to the star and do not exclude older or colder ejecta at
%substantially larger radii.  V509~Cas displayed variable CO first-overtone emission and absorption
%in 1979--1980, whereas the features were absent in spectra obtained
%in 1988, 2003, 2004, and 2019
%\citep[and references therein]{Kraus2023}.  The 2023 CHARA result provides a complementary spatial
%constraint: The CO-band brightness distribution is not measurably
%more extended than the adjacent continuum. The spectroscopic and
%interferometric measurements are therefore both consistent with the
%absence of a large, warm molecular atmosphere in the present
%post-heating state.

The compact continuum may indicate that the dense
pseudo-photosphere associated with the earlier evolutionary episode
has receded or become optically thin, exposing hotter atmospheric
layers. This interpretation is consistent with the historical
increase in the spectroscopically inferred effective temperature,
although the present observations do not uniquely determine the
physical mechanism responsible for that evolution.

\subsection{High luminosity does not guarantee a large CO atmosphere}
\label{sec:discussion_transition}

The current sample does not establish a universal temperature threshold for CO extension. The cool-supergiant literature covers a narrow $\teff$ range, YHG temperatures vary with phase, and the AMBER survey found no significant temperature correlation within its RSG sample \citep{ArroyoTorres2015}. 
Within the present heterogeneous sample, the CO extension appears to
depend on the atmospheric state rather than on a single sharply
defined temperature threshold.

Despite its high luminosity, V509~Cas lacks the strong CO-band
enlargement observed in the cooler and more dynamically active
hypergiants $\rho$~Cas and RW~Cep
(Figure~\ref{fig:ratio_comparison}). The revised distance and
luminosity reported by \citet{Kasikov2026} place V509~Cas near the
Humphreys--Davidson luminosity boundary, with
$\log(L/L_\odot)\simeq5.83$ \citep{Humphreys1979}. Its compact CO-forming atmosphere
 cannot be attributed to comparatively low luminosity.

The contrast among these objects suggests that luminosity alone does
not determine the molecular-atmosphere extension. Differences in
temperature, atmospheric density, pulsation phase, and recent shock, mass-loss history or presence of companion are also likely to be important.

%%%%%%%%%%%%%%%%%%%%%%%%%%%%%%%%%%%%%%%%%%%%%%%%%%%%%%%
\subsection{Angular-diameter evolution during the post-heating state}
\label{sec:discussion_diameter}

The most nearly like-for-like historical comparison uses the
PTI limb-darkened diameter derived with a Kurucz atmosphere correction,
$\theta_{\rm LD,PTI}=1.245\pm0.032$~mas
\citep{vanBelle2009}, and the CHARA fixed-Kurucz result,
$\theta_{\rm LD,CHARA}^{\rm Kurucz}=1.1862\pm0.0036$~mas.
The CHARA value is nominally smaller by $4.7\pm2.6$\%,
corresponding to only $1.8\sigma$. The two measurements are
statistically consistent.

The NExScI PTI archive\footnote{\url{https://nexsci.caltech.edu/software/PTISupport/pti_year.shtml}} shows that V509~Cas was observed on eight
nights: 2002 July 25 and 29, 2002 November 19, 2006 August 8, 19,
and 22, and 2006 September 3 and 4. These observations were obtained
after the rapid increase in the spectroscopically inferred effective
temperature, which had reached $\teff=7900\pm200$~K by the late
1990s \citep{Israelian1999}. The PTI--CHARA comparison 
does not directly measure any contraction associated with the earlier
heating episode. Instead, it tests whether the apparent photospheric
diameter continued to evolve during the subsequent post-heating state.

\begin{figure*}[t]
\centering
\safeincludegraphics[width=\textwidth]{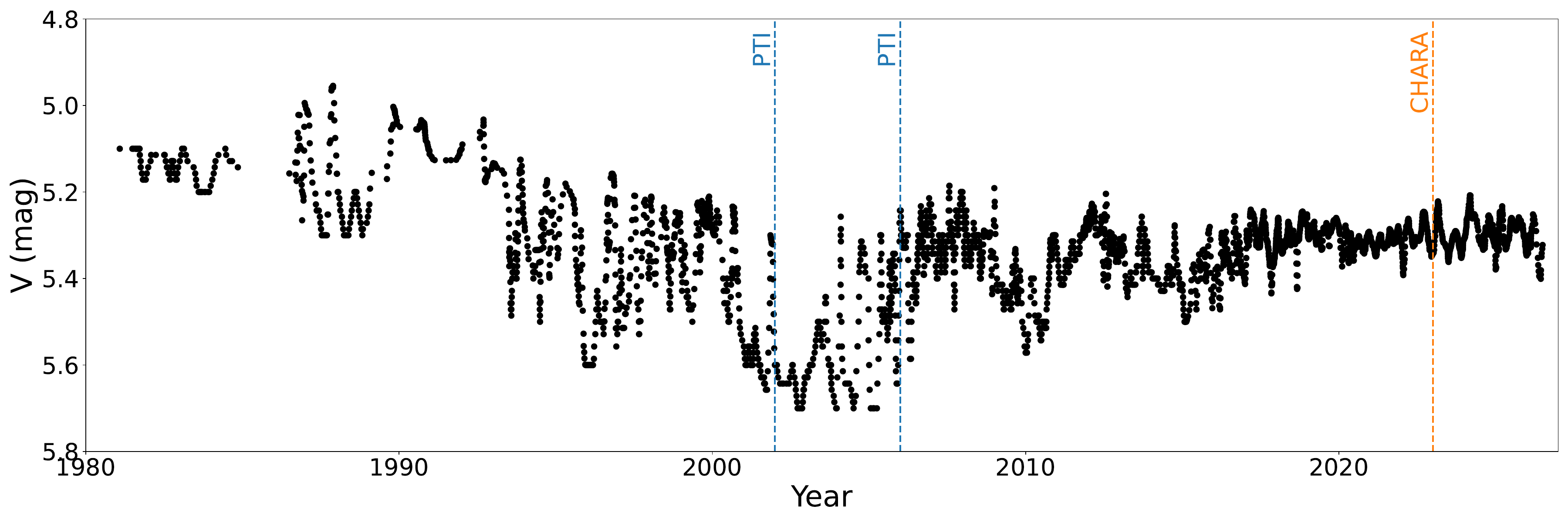}
\caption{
AAVSO $V$-band photometry \citep{Watson2006} of V509~Cas from 1980 through 2026.
The blue vertical lines mark the 2002 and 2006 PTI observing campaigns,
and the orange line marks the CHARA observation on 2023 October 28.
After the large secular variations of the late twentieth century, the
star entered a post-heating state with a comparatively stable
long-term mean brightness but persistent short-term variability,
likely associated with pulsation. The interferometric observations
 sampled the same broad evolutionary state, but potentially
different pulsation phases.
}
\label{fig:aavso}
\end{figure*}

Figure~\ref{fig:aavso} places the interferometric observations in
their long-term photometric context. At the 2002 PTI epoch, V509~Cas was up to approximately
0.3~mag fainter in the $V$ band than near the 2023 CHARA epoch,
whereas its brightness during the 2006 campaign was closer to the
2023 level. This difference represents a lower visual-band
flux, but it should not be interpreted directly as a change in
bolometric luminosity because the visual brightness of a yellow
hypergiant is sensitive to effective temperature, atmospheric opacity,
circumstellar extinction, and pulsation phase. The nominally larger
PTI diameter may reflect a different pulsational or
atmospheric state, potentially involving a larger and cooler apparent
photosphere.

The AAVSO light curve also shows that the large secular brightness
changes observed before approximately 2000 were followed by a
comparatively stable long-term mean, although substantial short-term
variability persisted. The PTI and CHARA observations 
sampled the same broad post-heating state, but not necessarily the
same pulsation phase. Their statistically consistent diameters show
no significant secular change in angular size over 17--21~yr, while
phase-dependent radius variations at the few-percent level remain
possible.

For completeness, JSDC2 lists
$\theta_{\rm UD,H}=1.5302$~mas and
$\theta_{\rm LD}=1.5665\pm0.12$~mas for V509~Cas
\citep{Bourges2017}. These are photometrically inferred catalog
estimates rather than direct interferometric measurements. The PTI
diameter is approximately 20\% smaller than the JSDC2 estimate.
\citet{Kasikov2026} argued that the photometric catalog diameter is
likely overestimated because the applicable surface-brightness
relations are poorly calibrated for variable yellow hypergiants.
We  do not treat the JSDC2 estimate as an additional
historical interferometric epoch.

%%%%%%%%%%%%%%%%%%%%%%%%%%%%%%%%%%%%%%%%%%%%%%%%%%%%%%%%
\subsection{Companions and the diversity of YHG states}
\label{sec:discussion_binarity}

The strongest evidence for a hot companion to V509~Cas comes
from the ultraviolet observations of \citet{SticklandHarmer1978}.
The short-wavelength IUE spectrum, covering approximately
1150--2000~\AA, resembled that of an early B-type star. A
spectral type of approximately B1~V was inferred from both the
ultraviolet energy distribution and the relative strengths of the
Si~IV and C~IV absorption features. At longer wavelengths, the 
contribution was higher, with the two
components estimated to contribute comparable flux near
2700~\AA\ and  approximately
3000~\AA. This wavelength dependence is important because
V509~Cas was cooler than 6000~K at the time of the original
observations and was expected to contribute relatively
little far-ultraviolet flux
\citep{Nieuwenhuijzen2012,Lobel2013}.

The circumstellar [N~II] emission provides supporting, although
indirect, evidence. These forbidden lines form in an extended,
low-density ionized environment that is difficult to maintain
using the radiation field of a cool yellow hypergiant.
\citet{SticklandHarmer1978} suggested that the proposed
B1~V star could ionize the circumstellar gas and power part of
the observed [N~II] and radio emission. The [N~II] emission has
remained remarkably stable during historical and subsequent spectroscopic
monitoring \citep{ShefferLambert1992,Klochkova2019}. This interpretation is
physically plausible, but it is not an independent dynamical
confirmation of the companion; shocks or ionization associated
with the previous mass-loss history may also contribute.

The orbital configuration remains essentially unconstrained.
\citet{SticklandHarmer1978} considered whether the historical
radial-velocity range could represent orbital motion. Under
assumed component masses, this produced an illustrative
separation of approximately 8.5~au and a period near 4~yr,
corresponding to about 2.5~mas at our adopted distance. However,
they concluded that the velocity variations more likely reflected
motions in the hypergiant atmosphere. In a different approach,
\citet{Piters1988} modeled the radio spectrum as emission from
the hypergiant wind photoionized by the B-type star. Their model
favored a separation near 200~au and a period of approximately
500~yr, corresponding to about 80~mas at the distance they
adopted, or approximately 59~mas at 3.4~kpc. Neither estimate
constitutes a measured orbit.

Long-term spectroscopy has not revealed a convincing coherent
orbital radial-velocity signal \citep{Klochkova2019}. Nevertheless,
the photospheric and atmospheric velocities vary by many
kilometers per second because of pulsation and atmospheric
motions, making a low-amplitude or very long-period orbital
signal difficult to isolate. The absence of such a signal 
does not by itself rule out the ultraviolet companion.

Both historical angular scales lie within our companion-search
region, and no significant asymmetric signal is detected in
either the MIRC-X $H$ band or the MYSTIC $K$ band.
 We estimated whether a normal B1~V companion
would have been detectable at our achieved contrasts.
Main-sequence calibrations give typical B1~V parameters of
$T_{\rm eff}\simeq25{,}000$--$27{,}000$~K and
$R\simeq5.5$--$6\,R_\odot$ \citep{Eker2018}. Adopting
these, together with
$T_{\rm eff}=7900$~K and $R\simeq433\,R_\odot$ for V509~Cas,
gives approximate continuum contrasts of
$\Delta H\simeq7.6$~mag and $\Delta K\simeq7.7$~mag.
These are fainter than our median $3\sigma$ limits of
$\Delta H=5.28$~mag and $\Delta K=6.48$~mag.
The CHARA observations do not exclude the B1~V
companion proposed from ultraviolet spectroscopy, but they do
exclude companions that are substantially brighter in the
near-infrared.

%Both historical angular scales lie within our companion-search region, and no significant asymmetric signal is detected in either the MIRC-X $H$ band or the MYSTIC $K$ band. However, the historical estimate of $\Delta V\simeq5$~mag does not determine the corresponding near-infrared contrast. The latter depends on the temperatures, radii, extinction, and evolutionary states adopted for both components. In particular, a hot main-sequence star may contribute appreciable far-ultraviolet flux while remaining extremely faint relative to the hypergiant in the near-infrared. Our non-detection therefore excludes companions brighter than the local $H$- and $K$-band injection--recovery limits, but does not exclude the proposed B1~V companion.

The recently detected companion candidate of Betelgeuse illustrates
the difficulty of detecting faint companions in continuum observations. Using VLT/SPHERE-ZIMPOL,
\citet{Montarges2026} detected Betelgeuse~B at a projected
separation of $52.32\pm0.18$~mas, with an optical continuum
flux ratio of $(8.24\pm1.04)\times10^{-4}$. This corresponds
to $\Delta m=7.71\pm0.14$~mag and is numerically fainter than the
$H$- and $K$-band contrast limits reached here, although the comparison
is not band-equivalent. The contrast was measured at 644.9~nm rather
than in the $H$ or $K$ band, and Betelgeuse~B is inferred to
be substantially later than the proposed B1~V companion of
V509~Cas. The comparison is illustrative rather
than directly quantitative.

In contrast, the proposed close companion of the yellow
hypergiant HR~5171~A represents a much brighter near-infrared
case. Interferometric modeling placed this component at a
separation of $1.45\pm0.07$~mas, projected against the limb of
the primary, and assigned it $12\pm3$\% of the total flux at
2.1~$\mu$m, corresponding to a quoted contrast of
$\Delta K\simeq2.3$~mag \citep{Chesneau2014}. A component
this bright would lie well above the contrast limits obtained
for V509~Cas. The Betelgeuse
and HR~5171~A cases illustrate the broad range of
companion contrasts encountered among evolved massive stars:
a luminous interacting component can be recovered
interferometrically, whereas a physically important but much
fainter companion can remain inaccessible to near-infrared
continuum observations.

High-spectral-resolution interferometry across the hydrogen
Br$\gamma$ line at 2.166~$\mu$m may provide a complementary
test if emission associated with the proposed companion or a
wind-interaction region is spatially displaced from the hypergiant.
In KQ~Pup, VLTI/GRAVITY differential phases across Br$\gamma$
traced compact emission associated with the hot subsystem and its
accretion flow, enabling relative astrometry even though the hot
components were too faint for continuum detection
\citep{Jadlovsky2025}.

\section{Conclusions and Future Prospects}
\label{sec:conclusions}

We presented simultaneous 2023 CHARA/MIRC-X and MYSTIC observations of V509~Cas. The main results are:

\begin{itemize}
\item The continuum photosphere is resolved with $\theta_{\rm UD,H}=1.148\pm0.003$~mas and $\theta_{\rm UD,K}=1.173\pm0.007$~mas. Fixed atmosphere CLVs give a joint $H+K$ diameter of 1.1840--1.1950~mas, with the SATLAS value $1.195\pm0.007$~mas adopted as the reference photospheric scale.
\item The 2.00--2.30~$\mu$m continuum and 2.30--2.37~$\mu$m CO diameters give $\mathcal{R}_{\rm CO}=1.011\pm0.004$, a marginal increase below the adopted $3\sigma$ detection criterion. For comparison with previously resolved molecular atmospheres, we adopt $\mathcal{R}_{\rm CO}=1.05$ as a conservative sensitivity threshold; no enlargement reaches that threshold.
\item The closure phases are consistent with a centrosymmetric brightness distribution, and no strong compact broadband continuum halo is required. Thus, spectroscopically visible atomic gas can remain while the near-infrared continuum and CO-forming regions are compact.
\item No companion is detected in either band. The median $3\sigma$
limits are $\Delta H=5.28$~mag and $\Delta K=6.48$~mag, while the
90\% completeness limits are $\Delta H=5.06$~mag and
$\Delta K=6.27$~mag. These are wavelength-dependent flux-ratio
constraints and do not exclude all forms of binarity.
\item The Kurucz-based PTI and CHARA limb-darkened diameters differ by only
$1.8\sigma$ and show no significant secular change in angular size.
The observations sampled different visual-brightness and likely
different pulsation states, so radius variations at the few-percent
level remain possible.
\end{itemize}

V509~Cas is a high-luminosity but molecularly compact YHG. Its contrast with the extended CO atmospheres of $\rho$~Cas and RW~Cep indicates that high luminosity alone does not guarantee a large molecular extension. Future same-band CHARA observations and contemporaneous spectroscopy can test whether the compact CO state persists and whether it changes with the atomic-line spectrum or photometric variability.

\begin{acknowledgments}
We thank the anonymous referee for helpful comments that improved the manuscript.
We thank Stephen Ridgway and Noelle Elmberg for reading the manuscript and for comments that improved the presentation of this work. We also thank the CHARA support staff and the MIRC-X/MYSTIC collaboration for their valuable contributions to the construction, maintenance, and operation of the instruments. 
This work is based upon observations obtained with the Georgia State University Center for High Angular Resolution Astronomy Array at Mount Wilson Observatory. The CHARA Array is supported by the National Science Foundation under Grant Nos. AST-2034336 and AST-2407956. Institutional support has been provided by the GSU College of Arts and Sciences, Office of the Provost, and Office of the Vice President for Research and Economic Development. SK acknowledges funding for MIRC-X from the European Research Council under the European Union's Horizon 2020 research and innovation programme (Starting Grant No. 639889 and Consolidated Grant No. 101003096). JDM acknowledges funding for the development of MIRC-X (NSF AST-2009489) and MYSTIC (NSF ATI-1506540 and NSF AST-1909165). We grate fully acknowledge the contributions of the AAVSO observer community, whose photometric data and metadata resources were used in this study and made available through the AAVSO’s scientific archives. ChatGPT  was used for language editing and readability improvements. The authors reviewed and verified all scientific content.
\end{acknowledgments}

\facilities{CHARA (MIRC-X, MYSTIC), AAVSO}

\software{PMOIRED, ExoTiC-LD}

\bibliography{references_v4}
\bibliographystyle{aasjournal}

\end{document}